\documentclass[12pt]{article}
\usepackage{color}
\usepackage{amsfonts,amssymb,amsmath}
\usepackage[export]{adjustbox}
\usepackage{graphicx}
\usepackage[T1]{fontenc}
\usepackage[numbers,sort&compress]{natbib}
\graphicspath{ {./images/} } \textheight 9in \textwidth  6.5in
\newcounter{tempeq}
\begin{document}
\title{\textsf{Enhancing the performance of an open quantum battery by adjusting its velocity}}
\author{B. Mojaveri\thanks{Email: bmojaveri@azaruniv.ac.ir; bmojaveri@gmail.com},
\hspace{2mm}R. Jafarzadeh Bahrbeig\thanks{Email:
r.jafarzadeh86@gmail.com},\hspace{2mm}M. A. Fasihi
\thanks{Email: ma-fasihi@azarunic.ac.ir}, and S. Babanzadeh\thanks{Email: s.babanzadeh@azaruniv.ac.ir}\\
{\small {Department of Physics, Azarbaijan Shahid Madani University,
PO Box 51745-406, Tabriz, Iran \,}}} \maketitle
\begin{abstract}
The performance of open quantum batteries (QBs) is severely limited
by decoherence due to the interaction with the surrounding
environment. So, protecting the charging processes against
decoherence is of great importance for realizing QBs. In this work
we address this issue by developing a charging process of a
qubit-based open QB composed of a qubit-battery and a qubit-charger,
where each qubit moves inside an independent cavity reservoir. Our
results show that, in both the Markovian and non-Markovian dynamics,
the charging characteristics, including the charging energy,
efficiency and ergotropy, regularly increase with increasing the
speed of charger and battery qubits. Interestingly, when the charger
and battery move with higher velocities, the initial energy of the
charger is completely transferred to the battery in the Markovian
dynamics. In this situation, it is possible to extract the total
stored energy as work for a long time. Our findings show that open
moving-qubit systems are robust and reliable QBs, thus making them a
promising candidate for experimental implementations.
\end{abstract}
\section{introduction}
In recent years, with advancements in quantum thermodynamics, there
has been a radical change of perspective in the framework of energy
manipulation based on the electrochemical principles. The
possibility to create an alternative and efficient energy storage
device at small scale introduces the concept of the quantum battery
(QB), which was proposed by Alicki and Fennes in the 2013's
\cite{Alicki}, and  subsequently became into a significant field of
research. As their name indicates, QBs are finite dimensional
quantum systems that are able to temporarily store energy in their
quantum degrees of freedom for later use. The fundamental strategy
for developing the idea of QBs is based on their non-classical
features such as quantum coherence, entanglement and many-body
collective behaviors that can be cleverly exploited to achieve more
efficient and faster charging processes than the macroscopic
counterpart \cite{Strasberg, Vinjanampathy, Goold, Campisi,
Gelbwaser, Horodecki}. A QB is charged based on an interaction
protocol between QB itself with either an external field or a
quantum system which serves as a charger. It is then discharged into
a consumption hub based on the same protocol. When the battery
enters into an interaction with the charger, it transitions from a
lower energy level into the higher ones and will be charged. So far,
a variety of powerful charging protocols have been proposed in
different platforms, including two-level systems \cite{Farin, Zhang,
Fus}, harmonic oscillators \cite{Cata}, and hybrid light-matter
systems \cite{Maze, Manzo, Cond}. Some proposals have been also
devoted to implement QBs based on the two-level systems such as
trapped ions \cite{Forn, Lv}, cold atoms \cite{Bau} and
superconducting qubits \cite{Devoret}.

 Due to the fact that a real quantum system inevitably interacts with
its environment, studying QBs from the open quantum systems
perspective is attracting considerable interest. The interaction of
a QB with its surrounding environments causes the leakage of the
coherence of battery to the environment, leading to decoherence
effect in the battery. Such an adverse effect often plays a negative
role in the charging and discharging performance of QBs \cite{Camp,
Farin1, Carega}. Decoherence brought during the charging process
tends to lead QBs to a non-active (passive) equilibrium state in
which work extracting from the QBs is often impossible \cite{Barra}
in a cyclic unitary process. The environmental-induced noises also
affect QBs that are disconnected from both charger and consumption
hub and cause self-discharging of that QBs \cite{San0, Pedro,
Salimi}. Therefore, designing a more robust battery against the
environmental dissipations is valuable step for implementation of
QBs in the real-life. Recently, researchers have devoted efforts not
only to studying the effect of the environment on QBs, but also to
exploit non-classical effect as well as to developing open system
protocols to stabilize the charging cycle performance through
quantum control techniques. For example, Kamin et al \cite{Kamin1}
studied the charging performance of a qubit-based QB charged by the
mediation of a non-Markovian environment. They revealed the
non-Markovian property is beneficial for improving charging cycle
performance. In Ref. \cite{Squeezing}, the authors studied dynamics
of a continuous variable QB coupled weakly to the squeezed thermal
reservoir and managed to control the performance of the charging
process by boosting the quantum squeezing of reservoir. A feasible
route for harnessing loss-free dark states for stabilizing the
stored energy of a qubit-based open QB has been introduced in
\cite{Dark}. In addition to the above considerations, several other
protocols have been developed to protect the charging cycle of QBs
such as feedback control method \cite{Mitch, Shao, Ios}, convergent
iterative algorithm \cite{Borhan}, Bang-Bang modulation of the
intensity of an external Hamiltonian \cite{Franc}, inhiring an
auxiliary quantum system \cite{Behzadi}, modulating the detuning
between system and reservoir \cite{Yu0}, stimulated Raman adiabatic
passage technique \cite{Baris}, engineering quantum environments
\cite{Segal}, etc.

 On the other hand, according to the previous studies on
the Markovian and non-Markovian dynamics of open two-qubit systems,
translational motion of qubits provides novel insights for
stabilizing qubit-qubit entanglement against the environmental
induced dissipations by suitably adjusting the velocities of the
qubits \cite{Epjp0, morteza0, Chao0, sare0, Golkar1, Epjp1, MPLA,
Wang00}. We want here to use this safeguard capability of the
motional properties to improve the charging cycle performance of the
open qubit-based QBs. For this end, we consider a moving-biparticle
system composed of a qubit-battery and a qubit-charger that
independently interacts with their local environments. The battery
qubit here is charged with the help of the dipole-dipole interaction
with the charger qubit. We will investigate how the translational
motion of qubits affects the charging process of QB. Our results
show that translational motion of qubits always plays a constructive
role in protecting QB from decay induced by the environment. This
work is organized as follows: in Sec. 2, we introduce and describe
several figures of merit for characterizing the performance of QBs.
In Sec. 3, we illustrate our model and obtain explicit expressions
for the reduced density matrix of the QB and the charger. In Sec. 4
we present the results of our numerical simulations in the context
of their physical significance. Finally, Sec. 5 concludes this
paper.
\section{Figures of Merit}
Let us consider a QB modeled as a quantum system with d-dimensional
Hilbert space $\mathcal{H}$ and Hamiltonian $H_B$ such that
\renewcommand\theequation{\arabic{tempeq}\alph{equation}}
\setcounter{equation}{-1}
\addtocounter{tempeq}{1}\begin{eqnarray}\label{Bat}
H_B=\sum_{i=1}^{d} \varepsilon_i
|\varepsilon_i\rangle\langle\varepsilon_i|,
\end{eqnarray}
with non-degenerate energy levels $\varepsilon_i \leq
\varepsilon_{i+1}$. Internal energy of QB is given by $Tr(\rho_B
H_B)$, where $\rho_B$ is the state of the battery. Charging a QB
means brings the quantum system from a lower energy state $\rho_B$
to a higher energy state $\rho_B^\prime$, while discharging refers
to the inverse process, i.e., brings the quantum system from a
higher energy state $\rho_B^\prime$ to a lower one
$\rho_B^{\prime\prime}$:
\renewcommand\theequation{\arabic{tempeq}\alph{equation}}
\setcounter{equation}{-1}
\addtocounter{tempeq}{1}\begin{eqnarray}\label{den}
\texttt{Tr}\left\{\left(\rho_B^\prime-\rho_B\right) H_B\right\}\geq0,\qquad\qquad\qquad\qquad charging \nonumber \\
\texttt{Tr}\left\{\left(\rho_B^{\prime\prime}-\rho_B^\prime\right)
H_B\right\}\geq0.\qquad\qquad\qquad\quad \;\;discharging
\end{eqnarray}
Therefore, in a charging process, the actual stored energy of QB at
time $t$, regarding the initial energy, can be expressed as follows
\cite{Alicki}
\renewcommand\theequation{\arabic{tempeq}\alph{equation}}
\setcounter{equation}{-1} \addtocounter{tempeq}{1}\begin{equation}
\Delta E_B=\texttt{Tr}\{\rho_B(t) H_B\}-\texttt{Tr}\{\rho_B(0)
H_B\}.
\end{equation}
A complete converting the stored energy into valuable work is
impossible without dissipation of heat according to the second law
of thermodynamics. The maximum amount of energy extracted from a
given quantum state $\rho_B=\sum_{i} r_i |r_i\rangle\langle r_i|$,
($ r_i \geq r_{i+1}$) through a cyclic unitary operation is called
ergotropy \cite{Allahverdyan}. This quantity can be defined as
\cite{Allahverdyan, Franc0, Cakmak0}
\renewcommand\theequation{\arabic{tempeq}\alph{equation}}
\setcounter{equation}{-1}
\addtocounter{tempeq}{1}\begin{equation}\label{ergo}
\mathcal{W}=\texttt{Tr}\{\rho_B
H_B\}-\texttt{min}_U\,\texttt{Tr}\{U\rho_B U^{\dagger} H_B\},
\end{equation}
where the minimization is taken over all possible unitary
transformations acting locally on such system. It has been shown in
\cite{Allahverdyan} that no work can be extracted from the passive
counterpart of $\rho_B$ with the form $\sigma_{\rho_B}=\sum_{i} r_i
|\varepsilon_i\rangle\langle\varepsilon_i|$. The unique unitary
transformation $U=\sum_i |\varepsilon_i\rangle\langle r_i|$ on the
$\rho$ minimizes $\texttt{Tr}(U\rho_B U^{\dagger} H_B)$, and when
inserted in Eq. (\ref{ergo}) yields the following expression for the
ergotropy
\renewcommand\theequation{\arabic{tempeq}\alph{equation}}
\setcounter{equation}{-1} \addtocounter{tempeq}{1}\begin{equation}
\mathcal{W}=\sum_{i,j} r_j \varepsilon_i\left(|\langle
r_j|\varepsilon_i\rangle|^2-\delta_{ij}\right).
\end{equation}
In order to quantify the amount of extractable energy, the
efficiency $\eta$ is defined as the ratio between the ergotropy
$\mathcal{W}$ and the total charging energy $\Delta E_B$
\renewcommand\theequation{\arabic{tempeq}\alph{equation}}
\setcounter{equation}{-1} \addtocounter{tempeq}{1}\begin{equation}
\eta=\frac{\mathcal{W}}{\Delta E_B}.
\end{equation}\begin{figure}
\centering
\includegraphics[keepaspectratio, width=.45\textwidth]{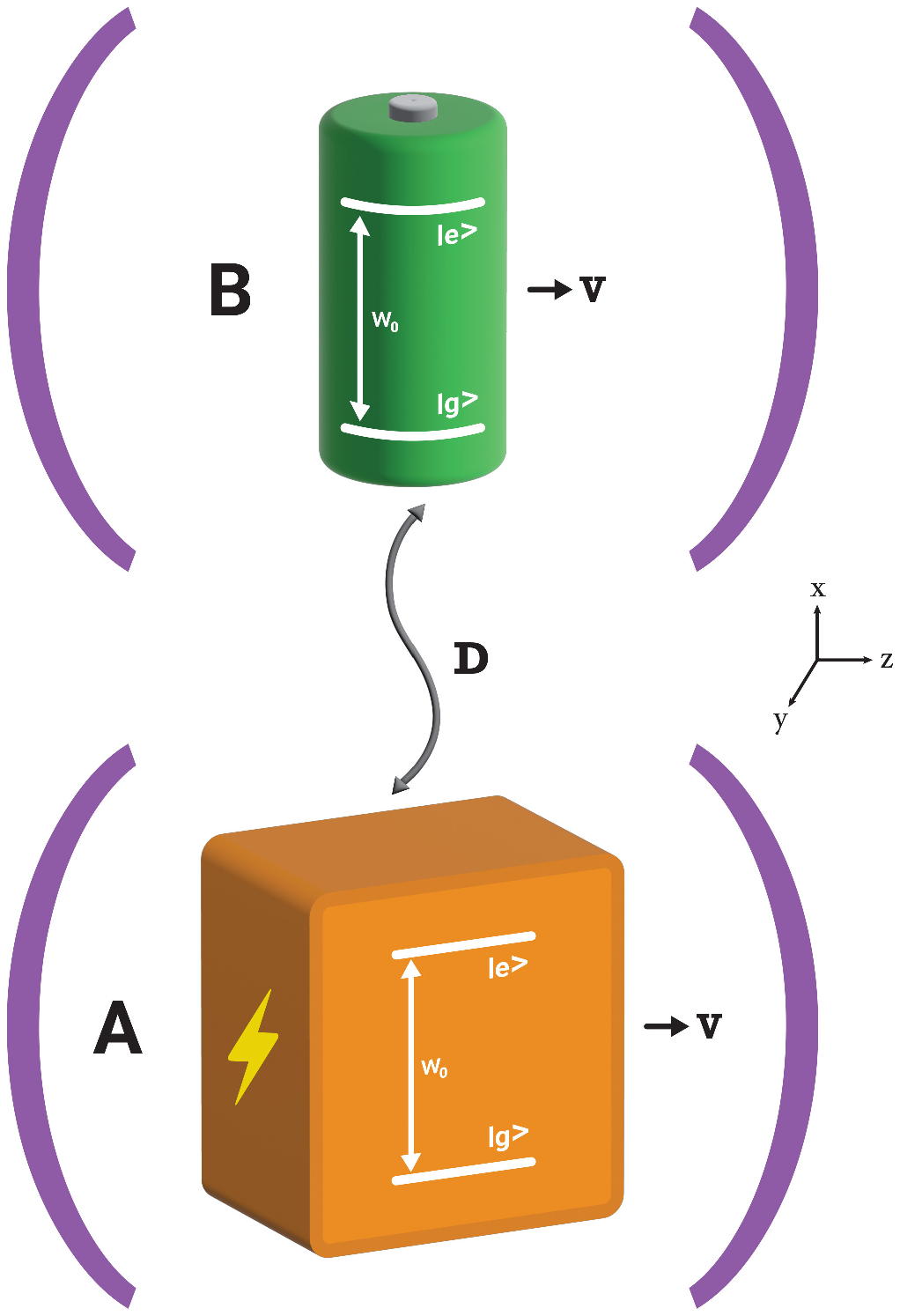}
\caption{Schematic illustration of a qubit-based open QB composed of
a qubit-battery and a qubit-charger moving along the z-axis of two
distinct but identical cavity reservoir. The qubits move with
constant speed $v$ and are also coupled to each other through the
dipole-dipole interaction.}
\end{figure}
\section{Open Moving-Quantum Battery}
The open QB under consideration is composed of an atomic two-qubit
system, the qubit $A$ as a charger and the qubit $B$ as a quantum
battery, coupled to each other trough the dipole-dipole interaction.
The battery and charger qubits coupled locally to two independent
zero-temperature cavity reservoirs (see Fig. 1). We assume that each
qubit moves along the $z$-axis of its cavity at a constant
non-relativistic speed $v$. For simplicity we neglect here any
scattering \cite{Engl} or trapping \cite{Haro} effects and consider
the translational motion of the atom qubits being classically. Under
the dipole and rotating wave approximation, the entire system is
ruled by Hamiltonian (setting $\hbar=1$)
\renewcommand\theequation{\arabic{tempeq}\alph{equation}}
\setcounter{equation}{-1} \addtocounter{tempeq}{1}\begin{equation}
H=H_0+H_{int},
\end{equation}
with
\renewcommand\theequation{\arabic{tempeq}\alph{equation}}
\setcounter{equation}{-1}
\addtocounter{tempeq}{1}\begin{eqnarray}\label{Ham}
&&\hspace{-1.15cm}
H_0=H_A+H_B+H_{R_A}+H_{R_B}=\sum_{j=A,B}\left(\frac{\omega_0}{2}
\sigma_{z}^{j} + \sum_{k}\omega_{k}^j a_{k}^{j\dag} a_{k}^j\right),\nonumber\\
&&\hspace{-1.2cm}H_{int}=H_{A-B}+H_{A-R_A}+H_{B-R_B}=D\left(\sigma_{+}^{A}\sigma_{-}^{B}+\sigma_{-}^{A}
\sigma_{+}^{B}\right) +\sum_{j=A,B}\sum_{k} f_k^j(z)
\left(\mathfrak{g}_{k}^j \sigma_{+}^{j} a^j_k +H.c.\right).
\end{eqnarray}
Here, H.c. stands for Hermitian conjugate, $\sigma_z^j$,
$\sigma_+^j$, and $\sigma_-^j$ $(j=A,B)$ are, respectively, the
population inversion, raising and lowering operators of the $j$th
qubit with transition frequency $\omega_0$. $a_k^{j\dagger}$ and
$a^j_k$ are, respectively, the creation and annihilation operators
of the $k$th mode of the cavity reservoir $j$ with the frequency
$\omega_k^j$. Also, $D$ is coupling constant of the dipole-dipole
interaction between the battery and charger qubits, and
$\mathfrak{g}_{k}^j$ is the coupling constant between the $j$th
qubit and $k$th mode of in the cavity reservoir $j$. The effect of
translation motion of the battery and charger qubits has been
included in the model by introducing the $z$-dependent shape
function $f_k^j(z)$ in the Hamiltonian $H_{int}$. When the battery
and charger qubits are moving with same constant velocity $v$, the
shape function $f_k^j(z=vt)$ can be taken into account as
\renewcommand\theequation{\arabic{tempeq}\alph{equation}}
\setcounter{equation}{-1} \addtocounter{tempeq}{1}\begin{equation}
f_k^j(z)=\sin[\omega_k^j(\beta t-\Gamma)],\qquad\qquad j=A,B
\end{equation}
where, $\Gamma=L/c$ with $L$ being the size of the cavity. Also,
$\beta=v/c$ where $c$ refers to the speed of light in the vacuum
space. This particular form of the shape function can be obtained by
imposing an appropriate boundary condition on the cavity reservoirs
\cite{Lenard, morteza0}. Here we describe the translational motion
of both battery and charger qubits by classical mechanics ($z=vt$).
To this end, we will choose the values of the parameters in such a
way that the de Broglie wavelength of qubit $\lambda_B$ is
significantly smaller than the wavelength $\lambda_0$ associated
with the resonant transition $\omega_0=\omega_n$ ($\omega_n$ is the
central frequency of the cavity field mode) \cite{mortezapour,
Cook}. Furthermore, we consider a situation in which the photon
momentum is relatively small than the atomic momentum and thus we
neglect the atomic recoil caused by the interaction with the
electric field \cite{Wilkens}. In the optical regime, to ignore the
atomic recoil and consider the translational motion of atoms as
classical, the velocity of qubits should be $v\gg 10^{-3}$
\cite{morteza0}.

In the interaction picture (IP) generated by the unitary
transformation $U=e^{-iH_0t}$, the Hamiltonian (\ref{Ham}) can be
written as follows
\renewcommand\theequation{\arabic{tempeq}\alph{equation}}
\setcounter{equation}{-1}
\addtocounter{tempeq}{1}\begin{eqnarray}\label{HIP}
&&\hspace{-1.5cm}H_{IP}=D\left(\sigma_{+}^{A}
\sigma_{-}^{B}+\sigma_{-}^{A} \sigma_{+}^{B}\right)+
\sum_{j=A,B}\sum_{k} f_k^j(z)\left(\mathfrak{g}_{k}^j \sigma_{+}^{j}
a_k^{j} e^{i(\omega_0-\omega_k^j)t}+\mathfrak{g}_k^{j \ast}
\sigma_{-}^{j}a_{k}^{j\dag} e^{-i(\omega_0-\omega_k^j)t}\right).
\end{eqnarray}
It is straightforward to show that the total excitation operator
$\hat{\mathcal{N}}=\sum_{j=A,B}\left(\sum_k\hat{a_k}^{j\dagger}\hat{a_k}^j+
\frac{1}{2}\hat{\sigma}_{z}^{j}\right)+1$, commutes with the total
Hamiltonian, i.e. $[H,\hat{\mathcal{N}}]=0$ and therefor it is the
constant of the motion. This allows us to decompose Hilbert space of
the entire qubit-cavity system,
$\mathcal{H}=\mathcal{H}_q\otimes\mathcal{H}_R$ spanned by the basis
$\{\left|i_A,j_B\right\rangle\otimes\left|n_1,n_2, ...,n_k,
...\right\rangle_{R_A}|_{n_1,n_2,...=0}^{\infty}
\otimes\left|m_1,m_2, ...,m_k,
...\right\rangle_{R_B}|_{m_1,m_2,...=0}^{\infty}\}$
$\left(i,j=e,g\right)$ into the excitation subspaces, as follows
\renewcommand\theequation{\arabic{tempeq}\alph{equation}}
\setcounter{equation}{-1} \addtocounter{tempeq}{1}
\begin{eqnarray}
&&\hspace{-14mm} \mathcal{H}=\oplus_{n=0}^{\infty} \mathcal{H}_{n}.
\end{eqnarray}
As a result of this decomposition, the dynamics of the entire
qubit-reservoir system can be restricted to the excitation subspaces
labeled by the total excitation number $n$. Here we are interested
to explore dynamics of the entire system in the single-excitation
subspace $\mathcal{H}_1$ spanned by vectors
$\{\left|g_A,g_B\right\rangle\otimes\left|1_k\right\rangle_{R_A}\left|0_k\right\rangle_{R_B}|_{k=0}^\infty,
\left|g_A,g_B\right\rangle\otimes\left|0_k\right\rangle_{R_A}\left|1_k\right\rangle_{R_B}|_{k=0}^\infty,
\left|e_A,g_B\right\rangle\otimes\left|0_k\right\rangle_{R_A}\left|0_k\right\rangle_{R_B},
\left|g_A,e_B\right\rangle\otimes\left|0_k\right\rangle_{R_A}\left|0_k\right\rangle_{R_B}\}$
in which the single excitation is either in one of the qubits or in
the k-th mode of one of cavity reservoirs. We consider a normalized
initial state of entire qubit-reservoir as a superposition of
$\left|e_A,g_B\right\rangle\left|0_k\right\rangle_{R_A}\left|0_k\right\rangle_{R_B}$
and
$\left|g_A,e_B\right\rangle\left|0_k\right\rangle_{R_A}\left|0_k\right\rangle_{R_B}$
states with the following form
\renewcommand\theequation{\arabic{tempeq}\alph{equation}}
\setcounter{equation}{-1}
\addtocounter{tempeq}{1}\begin{eqnarray}\label{sai0}
|\Psi(0)\rangle=\big[c_1(0) |e_{A},g_{B}\rangle +c_2(0)
|g_{A},e_{B}\rangle\big]\otimes |0\rangle_{R_A}|0\rangle_{R_B}.
\end{eqnarray}
For times $t>0$, we expand the state vector $|\Psi(t)\rangle$ in
terms of the vector basis of the single-excitation subspace
$\mathcal{H}_1$ as
\renewcommand\theequation{\arabic{tempeq}\alph{equation}}
\setcounter{equation}{-1}
\addtocounter{tempeq}{1}{\footnotesize\begin{eqnarray}\label{sai}
&&\hspace{-3.5cm}\left|\Psi(t)\right\rangle=\big[c_1(t)\left |e_{A},
g_B\right\rangle +c_2(t) \left|g_A, e_B\right\rangle\big] \otimes
\left|0_k\right\rangle_{R_A}\left|0_k\right\rangle_{R_B}
\nonumber\\
&&\hspace{-2.35cm}+\left|g_A, g_B\right\rangle\otimes\sum_{k}
\big[d_{k}(t)\left|1_k\right\rangle_{R_A}\left|0_k\right\rangle_{R_B}+d_{k}^{\prime}(t)
\left|0_k\right\rangle_{R_A}\left|1_k\right\rangle_{R_B}\big],
\end{eqnarray}}
where the time-dependent amplitudes satisfy the normalization
requirement
\renewcommand\theequation{\arabic{tempeq}\alph{equation}}
\setcounter{equation}{-1} \addtocounter{tempeq}{1}\begin{eqnarray}
\sum_{i=1}^2|c_i(t)|^2+\sum_k(|d_{k}(t)|^2+|d_{k}^{\prime}(t)|^2)=1.
\end{eqnarray}
By taking the partial traces over the field modes and subsystem A
(B), the reduced time-dependent density operator for the battery
(charger) in the $\{\left|e\right\rangle, \left|g\right\rangle\}$
basis is obtained as
\renewcommand\theequation{\arabic{tempeq}\alph{equation}}
\setcounter{equation}{0} \addtocounter{tempeq}{1}\begin{eqnarray}
&&\hspace{-2cm}\rho_A(t)=|c_1(t)|^2\left|e_A\right\rangle\left\langle
e_A\right|-\left(1-|c_1(t)|^2\right)\left|g_A\right\rangle\left\langle
g_A\right|\label{rob2},\\
&&\hspace{-2cm}\rho_B(t)=|c_2(t)|^2\left|e_B\right\rangle\left\langle
e_B\right|-\left(1-|c_2(t)|^2\right)\left|g_B\right\rangle\left\langle
g_B\right|\label{rob1}.
\end{eqnarray}

 Inserting Eq. (\ref{sai}) into the time dependent Schr\"{o}dinger
equation $H_{IP}|\Psi(t)\rangle=i\frac{d}{d t}|\Psi(t)\rangle$, with
$H_{IP}$ given in (\ref{HIP}), leads to the following set of
differential equations for time-dependent amplitudes
\renewcommand\theequation{\arabic{tempeq}\alph{equation}}
\setcounter{equation}{0} \addtocounter{tempeq}{1}\begin{eqnarray}
&&\hspace{-4cm}i\dot{c_1}(t)=D c_2(t)+\sum_{k} \mathfrak{g}_{k}^A
f_k^A(z)d_{k}(t)e^{i(\omega_0-\omega_{k}^A)}\label{c1t},\\
&&\hspace{-4cm}i\dot{c_2}(t)= D c_1(t)+ \sum_{k} \mathfrak{g}_{k}^B
f_k^B(z)d_{k}^{\prime}(t)e^{i(\omega_0-\omega_{k}^B)}\label{c2t},\\
&&\hspace{-4cm}i\dot{d}_{k}(t)=\mathfrak{g}_k^{A\ast}f_k^A(z)
c_1(t)e^{-i(\omega_0-\omega_{k}^A)t},\label{d1t}\\
&&\hspace{-4cm}i\dot{d}_{k}^{\prime}(t)=
\mathfrak{g}_k^{B\ast}f_k^B(z)
c_2(t)e^{-i(\omega_0-\omega_{k}^B)t}\label{d2t}.
\end{eqnarray}
By integrating Eqs. (\ref{d1t}) and (\ref{d2t}) with the initial
condition $d_{k}(0)=0$ and $d_{k}^{\prime}(0)=0$ and putting their
solutions, respectively, in Eqs. (\ref{c1t}) and (\ref{c2t}), we get
the following integro-differential equations for the amplitudes
$c_1(t)$ and $c_2(t)$
\renewcommand\theequation{\arabic{tempeq}\alph{equation}}
\setcounter{equation}{0} \addtocounter{tempeq}{1}\begin{eqnarray}
&&\hspace{-2cm}\dot{c_1}(t)=-iDc_2(t)+\int_{0}^{t}F_A(t-t^\prime)c_1(t^\prime)dt^\prime,\label{mt}\\
&&\hspace{-2cm}\dot{c_2}(t)=-iDc_1(t)+\int_{0}^{t}F_B(t-t^\prime)c_2(t^\prime)dt^\prime,\label{nt}
\end{eqnarray}
where
\renewcommand\theequation{\arabic{tempeq}\alph{equation}}
\setcounter{equation}{0} \addtocounter{tempeq}{1}\begin{eqnarray}
&&\hspace{-2cm}F_{A}(t-t^\prime)=\sum_{k} |\mathfrak{g}_{k}^A|^2
e^{i(\omega_0-\omega_{k}^A)(t-t^\prime)}\sin[\omega_k^A(\beta^A
t-\Gamma)]\sin[\omega_k^A(\beta^A t^\prime-\Gamma)],\\
&&\hspace{-2cm}F_{B}(t-t^\prime)=\sum_{k} |\mathfrak{g}_{k}^B| ^2
e^{i(\omega_0-\omega_{k}^B)(t-t^\prime)}\sin[\omega_k^B(\beta^B
t-\Gamma)]\sin[\omega_k^B(\beta^B t^\prime-\Gamma)],
\end{eqnarray}
are the memory correlation function of the reservoirs $A$ and $B$,
respectively. For simplicity, we suppose
$F_{A}(t-t^\prime)=F_{B}(t-t^\prime)=F(t-t^\prime)$. In the limit of
a large number of modes ( in the continuum limit ), the correlation
function $F(t-t^\prime)$ takes the following form
\renewcommand\theequation{\arabic{tempeq}\alph{equation}}
\setcounter{equation}{-1}
\addtocounter{tempeq}{1}\begin{equation}\label{kernel}
F(t-t^\prime)=\int d\omega J(\omega)
e^{i(\omega_0-\omega)(t-t^\prime)}\sin[\omega(\beta
t-\Gamma)]\sin[\omega(\beta t^\prime-\Gamma)],
\end{equation}
in which $J(\omega)$ is the spectral density of the cavity
reservoirs and has the Lorentzian form \cite{Lenard, Breuer0}
\renewcommand\theequation{\arabic{tempeq}\alph{equation}}
\setcounter{equation}{-1}
\addtocounter{tempeq}{1}\begin{equation}\label{lorentz}
J(\omega)=\frac{1}{2\pi}\frac{\gamma\lambda^2}{(\omega_0-\omega-\Delta)^2+\lambda^2},
\end{equation}
where $\lambda$ defines the spectral width of the coupling which is
connected to the memory time $\tau_E$ by the relation
$\tau_E=\lambda^{-1}$ and $\gamma$ refers to the qubit-environment
coupling strength which is related to the relaxation time scale
$\tau_R$ by $\tau_R \approx \gamma^{-1}$. Also $\Delta$ is the
detuning of $\omega_0$ and the central frequency of the cavity. The
weak and strong coupling regimes can be distinguished by comparing
$\tau_E$ and  $\tau_R$, in other words with an increasing
$\frac{\tau_E}{\tau_R}=\frac{\gamma}{\lambda}$ ratio, the
interaction will transition into a strong coupling or a non-Markovian regime \cite{Breuer0}.\\
By inserting the Eq. (\ref{lorentz}) into the Eq. (\ref{kernel}) and
after some calculations, in the continuum limit ($\Gamma \rightarrow
\infty$), the correlation function is simplified as
\renewcommand\theequation{\arabic{tempeq}\alph{equation}}
\setcounter{equation}{-1}
\addtocounter{tempeq}{1}\begin{equation}\label{ft}
F(t-t^\prime)=\frac{\gamma \lambda}{4} \cosh[\beta
\overline{\lambda}(t-t^\prime)] e^{-(\lambda-i\Delta) |t-t^\prime|}
\end{equation}
with $\overline{\lambda}=\lambda+i(\omega_0-\Delta)$.\\
In view of (\ref{ft}), taking the Laplace transformations of both
sides of the differential Eqs. (\ref{mt}) and (\ref{nt}) and using
the convolution property
$\mathcal{L}[\int_{0}^{t}\mathbf{A}(t-t^\prime) \mathbf{B}(t^\prime)
dt^\prime]=\mathbf{A}(s)\mathbf{B}(s)$ yields
\renewcommand\theequation{\arabic{tempeq}\alph{equation}}
\setcounter{equation}{0} \addtocounter{tempeq}{1}\begin{eqnarray}
&&\hspace{-2cm}sc_1(s)-c_1(0)=-iDc_2(s)-F(s)c_1(s),\label{ms}\\
&&\hspace{-2cm}sc_2(s)-c_2(0)=-iDc_1(s)-F(s)c_2(s),\label{ns}
\end{eqnarray}
where the functions $c_1(s)$ and $c_2(s)$ are the Laplace
transformations of the $c_1(t)$ and $c_2(t)$, respectively, and
$F(s)$ is the Laplace transforms of $F(t-t^\prime)$ which has the
following explicit form
\renewcommand\theequation{\arabic{tempeq}\alph{equation}}
\setcounter{equation}{-1} \addtocounter{tempeq}{1}\begin{eqnarray}
F(s)=\frac{\gamma\lambda}{4}\frac{s+\overline{\lambda}}{(s+\overline{\lambda})^2-\beta^2\overline{\lambda}\,^2}.
\end{eqnarray}
By reformulating the Eqs. (\ref{ms}) and (\ref{ns}), we get a
general solution for $c_1(s)$ and $c_2(s)$ as follows
\renewcommand\theequation{\arabic{tempeq}\alph{equation}}
\setcounter{equation}{0} \addtocounter{tempeq}{1}\begin{eqnarray}
&&\hspace{-2cm}c_1(s)=\frac{s+F(s)}{\big(s+F(s)\big)^2+D^2}c_1(0)-i\frac{D}{(s+F(s))^2+D^2}c_2(0),\\
&&\hspace{-2cm}c_2(s)=\frac{s+F(s)}{\big(s+F(s)\big)^2+D^2}c_2(0)-i\frac{D}{(s+F(s))^2+D^2}c_1(0).
\end{eqnarray}
In continuation, by applying the inverse Laplace transformation on
the both side of the above equations, we obtain finally $c_1(t)$ and
$c_2(t)$, as
\renewcommand\theequation{\arabic{tempeq}\alph{equation}}
\setcounter{equation}{0} \addtocounter{tempeq}{1}\begin{eqnarray}
&&\hspace{-2cm}c_1(t)=\frac{1}{2}\bigg(c_1(0)\Re(\mathcal{M}(t))-ic_2(0)\Im(\mathcal{M}(t))\bigg)\label{ct12},\\
&&\hspace{-2cm}c_2(t)=\frac{1}{2}\bigg(c_2(0)\Re(\mathcal{M}(t))-ic_1(0)\Im(\mathcal{M}(t))\bigg)\label{ct122},
\end{eqnarray}
where, $\Re(x)$ ($\Im(x)$) is real (imaginary) part of $x$, and
\renewcommand\theequation{\arabic{tempeq}\alph{equation}}
\setcounter{equation}{-1} \addtocounter{tempeq}{1}\begin{equation}
\mathcal{M}(t)=\sum_{i,j,k=1}^3\varepsilon_{ijk}\frac{ e^{q_it}
(q_j-q_k)\bigg((q_i+\overline{\lambda})^2-\beta
^2\overline{\lambda}^2\bigg)}{\prod_{i=1}^{3}\prod_{j=i+1}^{3}(q_i-q_j)},
\end{equation}
with $\varepsilon_{ijk}$ is the Levi-Civita symbol and $q_i (i=  1,
2, 3)$ are the roots of
\renewcommand\theequation{\arabic{tempeq}\alph{equation}}
\setcounter{equation}{-1} \addtocounter{tempeq}{1}\begin{equation}
q^3+q^2(2 \overline{\lambda}-i \text{D} )+q \left(\frac{\gamma
\lambda }{4}+\overline{\lambda} (\overline{\lambda}-2 i
\text{D})-\beta ^2\overline{\lambda}^2\right)+\frac{\gamma  \lambda
\overline{\lambda}}{4}+i \text{D} \overline{\lambda}^2\left(\beta
^2-1\right)=0.
\end{equation}

 With substitution (\ref{ct12}) and (\ref{ct122}), respectively, into the reduced density matrices
(\ref{rob1}) and (\ref{rob2}), and then using the $\Delta
E_{A(B)}=\texttt{Tr}\{\rho_{A(B)}(t)
H_{A(B)}\}-\texttt{Tr}\{\rho_{A(B)}(0) H_{A(B)}\}$ , the internal
energy of the charger and battery are deduced as
\renewcommand\theequation{\arabic{tempeq}\alph{equation}}
\setcounter{equation}{-1} \addtocounter{tempeq}{1}\begin{equation}
\Delta
E_A=\omega_0\left(|c_1(t)|^2-|c_1(0)|^2\right),\quad\quad\Delta
E_B=\omega_0\left(|c_2(t)|^2-|c_2(0)|^2\right).
\end{equation}
On the other hand, one can obtain ergotropy of the battery by
substitution Eq. (\ref{rob1}) with Eq. (\ref{ergo}). So, we have
\renewcommand\theequation{\arabic{tempeq}\alph{equation}}
\setcounter{equation}{-1} \addtocounter{tempeq}{1}\begin{equation}
 W_B=\omega_0\left(2|c_2(t)|^2-1\right)\Theta
\left(|c_2(t)|^2-\frac{1}{2}\right),
\end{equation}
where $\Theta(x-x_0)$ is the Heaviside function, which satisfies
$\Theta(x-x_0)=0$ for $x<x_0$, $\Theta(x-x_0)=\frac{1}{2}$ for
$x=x_0$ and $\Theta(x-x_0)=1$ for $x>x_0$.
\begin{figure}
\centering
\includegraphics[keepaspectratio, width=1\textwidth]{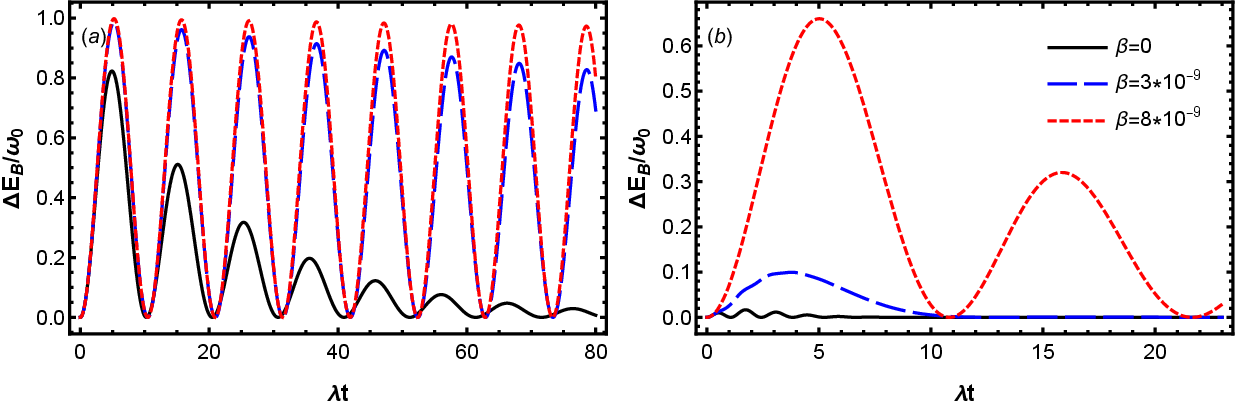}
\caption{Dynamics of the stored energy $\Delta E_B(t)$ for the
different values of $\beta$ by setting
$\omega_0=1.5\times10^9\lambda$, $D=0.3\lambda$ and
$\Delta=0.3\lambda$. The panels (a) displays the Markovian dynamics
with $\gamma=0.1 \lambda$, while the panels (b) displays the
non-Markovian dynamic with $\gamma=20\lambda$.}
\end{figure}
\begin{figure}
\centering
\includegraphics[keepaspectratio, width=1\textwidth]{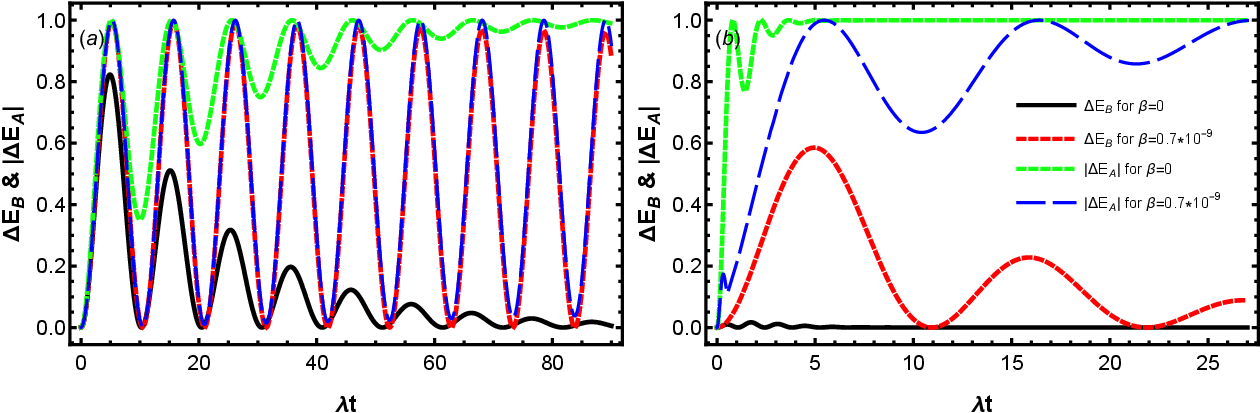}
\caption{Dynamics of the stored energy $\Delta E_B$ and internal
energy of charger $|\Delta E_A|$ for the different values of $\beta$
by setting $\omega_0=1.5\times10^9\lambda$, $D=0.3\lambda$ and
$\Delta=0$. The panels (a) displays the Markovian dynamics with
$\gamma=0.1 \lambda$, while the panels (b) displays the
non-Markovian dynamic with $\gamma=20\lambda$.}
\end{figure}
\section{Numerical Results and Discussion}
In this section, we will analyze the charging dynamics of the
introduced open moving-battery in the weak and strong coupling
regimes. In particular, we explore the role of the movement of QB on
the dynamical behavior of performance indicators including stored
energy, ergotropy and efficiency. In our following analysis, we
choose the optical regime parameters \cite{Hood, Pinkse} and
consider that qubit transition frequency as
$\omega_0=1.5\times10^{9}\lambda$. In what follows, we consider an
initial condition in which the battery is initially empty and the
charger has the maximum energy, i.e. $c_1(0)=0$, $c_2(0)=1$.
\begin{figure} \centering
\includegraphics[keepaspectratio, width=1\textwidth]{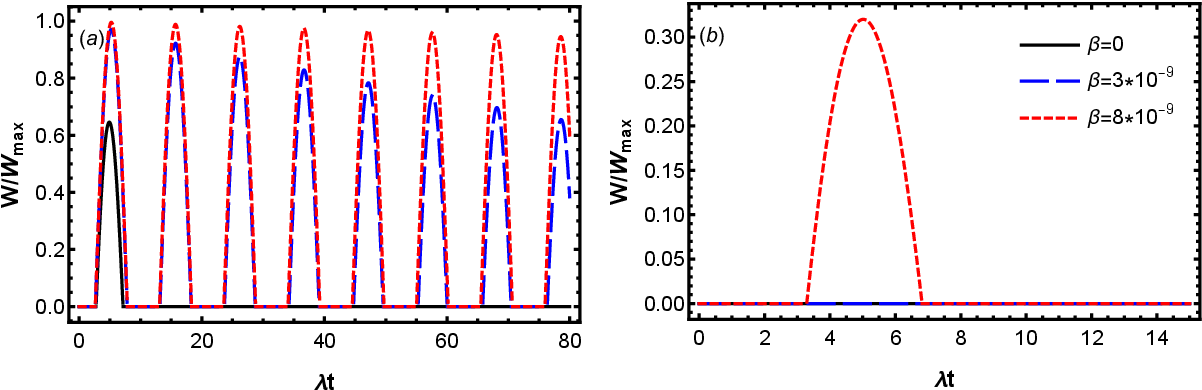}
\caption{Dynamics of ergotropy $W_B$ for the different values of
$\beta$ by setting $\omega_0=1.5\times10^9\lambda$, $D=0.3\lambda$
and $\Delta=0$. The panels (a) displays the Markovian dynamics with
$\gamma=0.1 \lambda$, while the panels (b) displays the
non-Markovian dynamic with $\gamma=20\lambda$.}\label{erg}
\end{figure}

 In Fig. 2, we plot the Markovian and non-Markovian dynamics of the stored energy $\Delta E_B$
for the initial state
$\left|\Psi(0)\right\rangle=\left|g\right\rangle_{A}\left|e\right\rangle_{B}\otimes
\left|0\right\rangle_{R_B}\left|0\right\rangle_{R_B}$, by
considering different values of the QB speed $\beta$. In panel (a),
the battery is charged in the Markovian dynamics with
$(\gamma=0.1\lambda)$, while in panel (b), it is charged in a
non-Markovian dynamics with $(\gamma=20\lambda)$. Here we consider a
situation at which the charger and battery's qubits are both in
resonance with the reservoir modes by setting $\Delta=0$. According
to this figure, the positive impact of the translational motion of
the charger and batter's qubits in controlling the stored energy of
battery is clearly visible in both Markovian and non-Markovian
charging processes. As can be seen in both Figs. 2(a) and (b), when
the charger and battery's qubits are at rest inside their cavity
reservoirs, the stored energy in the battery $\Delta E_B$ decays
into zero at sufficiently long times. However the rate of these
decays decreases regularly by gradual growth of the qubit velocity,
and therefore the energy stored in the battery and consequently the
charging process is strongly protected from the environmental
noises. Comparing Fig. 2(a) with Fig. 2(b) clearly reveals a
fundamental difference between Markovian and non-Markovian charging
processes. The maximal amount of stored energy in the Markovian
charging process is more than those of the non-Markovian charging
process. The reason stems from the nature of the qubit-cavity
coupling. In the non-Markovian charging process, the coupling
strength of charger's qubit to the cavity modes is greater than its
coupling to the battery's qubit, therefore, the initial internal
energy of charger has more tendency to evolve toward the reservoir
than to the battery. Moreover, since the motional effect of QB has
been included in battery-cavity and charger-cavity coupling
strength, it seems that increasing speed of QB decreases the
charger-cavity coupling strength in favor of to charger-battery
coupling strength, which increases the energy stored in the battery.

In order to get more insight to this area and a deeper understanding
of the relationship between the charger and battery energy, in Fig.
2 we have illustrated the energy stored in the battery at the end of
charging process as well as the energy that the charger loses at the
same time. Here $\Delta E_B$ and $|\Delta E_A|$ have been plotted as
a function of the dimensionless time $\lambda t$ for the qubit
velocities $\beta=0$ and $\beta=0.7\times 10^{-9}$ in the Markovian
and non-Markovian regimes. In the non-Markovian charging process,
$|\Delta E_A|$ is much more than $\Delta E_B$ for a given $\beta$ as
shown in Fig. 3(b). This implies that the internal energy of the
charger is not completely transferred to the battery. Fig. 3(b) also
shows that, when the charger and battery's qubits are at rest inside
their cavity reservoirs, the charger's qubit immediately loses a
large amount of its initial energy without being transferred to the
battery. However, increasing the qubit velocity (decreasing the
ratio of charger-cavity coupling strength to charger-battery
coupling strength) during the non-Markovian process, decreases the
initial loss-rate of the charger, and therefore improves the energy
transfer in the charging processes.

The relationship between the charger and battery energy in the
Markovian charging process is drastically different from that in the
non-Markovian charging process. One can infer from Fig. 3(a) that,
for the static battery-charger system ($\beta=0$), the total energy
of the charger can be transferred to the battery in the Markovian
short-charging process, where we have $|\Delta E_A|=\Delta E_B$.
Interestingly, when the qubits move with the velocity
$\beta=0.7\times10^{-9}$, $|\Delta E_A|=\Delta E_B$ holds at any
charging time. So, we conclude again that a robust Markovian
charging against the arisen dissipation can be achieved, when the
qubits move with higher velocities.
\begin{figure}
\centering
\includegraphics[keepaspectratio, width=1\textwidth]{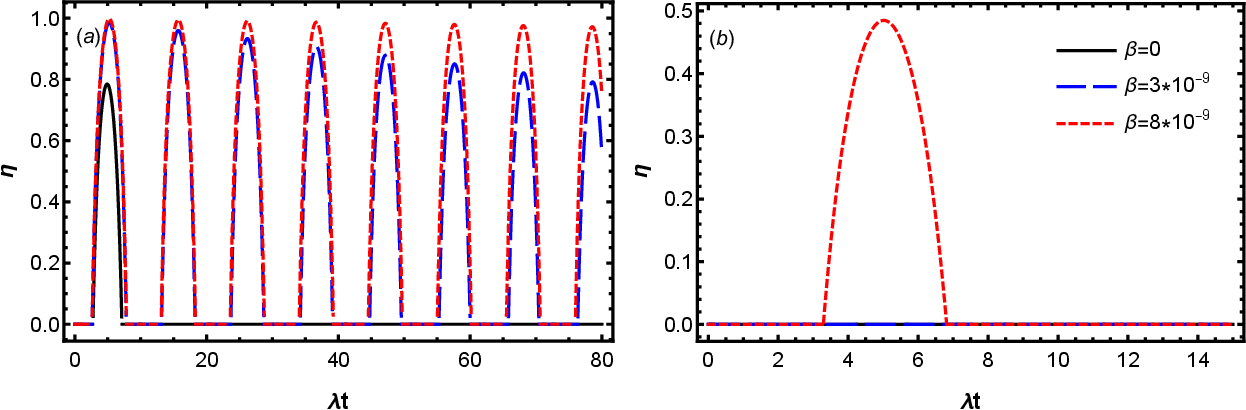}
\caption{Dynamics of efficiency $\eta(t)$ for the different values
of $\beta$ by setting $\omega_0=1.5\times10^9\lambda$,
$D=0.3\lambda$ and $\Delta=0$. The panels (a) displays the Markovian
dynamics with $\gamma=0.1 \lambda$, while the panels (b) displays
the non-Markovian dynamic with $\gamma=20\lambda$.}
\end{figure}

 In the following, we examine the influence of translational motion
of the battery-charger system on the dynamics of ergotropy. In Fig.
4, we plot $W/W_{max}$ as a function of $\lambda t$ for the
different values of $\beta$ in the Markovian (Fig. 4(a)) and
non-Markovian (Fig. 4(b)) regimes. Our numerical results in Fig.
4(a) and (b) illustrate that, the effect of translational motion of
QB on the ergotropy is also constructive in both Markovian and
non-Markovian regimes. Fig. 4(b) shows that, in the non-Markovian
regime, in the cases of stationary ($\beta=0$) and slowly moving
($\beta=3\times10^{-9}$) qubits, we are not able to extract useful
work from the QB, but in this regime a considerable work can be
extracted, as the qubits move with a higher velocity
($\beta=0.8\times10^{-9}$). Our numerical results in Fig. 4(a)
illustrate that, the effect of translational motion of QB on the
ergotropy is more considerable in the Markovian case. We observe
that, in the Markovian regime, increasing the speed of QB $\beta$
(decreasing the qubit-reservoir coupling) not only boosts the
ergotropy, but also increases the number of time zones in which work
can be extracted. Accordingly, a strong robust charging process can
be established in the higher speed limit, in which the extractable
work approaches to its maximum value.

 Finally, we examine the effect of translational motion
of QB on the Markovian and non-Markovian charging efficiency. The
results for Markovian and non-Markovian charging processes are
presented in Fig. 5(a) and 5(b), respectively. Here we consider the
same parameter values as Fig. 4. Comparing Figs. 4 and 3 reveals
that both ergotropy and efficiency are positively affected by the
translational motion of QB. However the efficiency is influenced
more than the ergotropy; the amount of increment in efficiency is
more than the ergotropy in both Markovian and non-Markovian charging
processes.
\section{Outlook and summary}
To summarize, we proposed a mechanism for robust charging process of
an open qubit-based quantum battery (QB) whose robustness can be
well controlled by the translational motion of the charger and
battery in both Markovian and non-Markovian dynamical regimes. Both
the battery and charger's qubits move with a same speed inside two
separated identical environments, and are directly coupled by the
dipole-dipole interaction. We showed that the stored energy,
ergotropy and efficiency of the moving QB regularly increased with
the gradual growth of the charger and battery speed, thereby
improving its charging performance. The constructive role of the
translational movement of QB in controlling the charging process
arises from the attachment of qubits velocities to the
qubit-reservoir coupling strength (see Eq. (\ref{Ham})). According
to the adopted charging protocol, a weak qubit-reservoir coupling is
required for a strongly robust charging process which can be
fulfilled by adjusting $\beta$ to the higher velocities.

 Our results represent a further control strategy to have a robust QB with
a natural implementation in cavity-QED context. The strategy can be
easily implemented also in the circuit-QED setups where the qubit
position slowly varies linearly with time and also the qubit-cavity
interaction is tuned through a sinusoidal position-dependent
coupling \cite{Jones}.

  In perspective, we believe that this strategy can be used
to control the performance of the discharging of a qubit-based QB to
an available consumption hub. Further efforts in this field can be
devoted to use the proposed strategy for improving the performance
of the two-photon based charging process where the moving-QB is
coupled with a cavity reservoir by means of a two-photon
relaxation.\\\\
\textbf{\large{Data availability}}\\ The datasets used and analysed
during the current study available from the corresponding author on
reasonable request.


\begin{thebibliography}{99}
\bibitem{Alicki} R. Alicki and M. Fannes, Entanglement boost for extractable work from ensembles of quantum batteries, Phys. Rev. E 87, 042123 (2013).
\bibitem{Strasberg} P. Strasberg, G. Schaller, T. Brandes, and M. Esposito, Quantum and information thermodynamics: A unifying framework based on repeated interactions, Phys. Rev. X 7, 021003 (2016).
\bibitem{Vinjanampathy} S. Vinjanampathy and J. Anders, Quantum thermodynamics, Cont. Phy. 57, 545 (2016).
\bibitem{Goold} J. Goold, M. Huber, A. Riera, L. del Rio, and P. Skrzypczyk, The role of quantum information in thermodynamics: a topical review, J. Phys. A: Math. Theor. 49, 143001 (2016).
\bibitem{Campisi} M. Campisi, P. H\"{a}nggi, and P. Talkner, Colloquium: Quantum fluctuation relations: Foundations and applications, Rev. Mod. Phys. 83, 1653 (2011).
\bibitem{Gelbwaser} D. Gelbwaser-Klimovsky, W. Niedenzu and G. Kurizki, Thermodynamics of quantum systems under dynamical control, Adv. At. Mol. Opt. Phys., 64, 329 (2015).
\bibitem{Horodecki} M. Horodecki and J. Oppenheim,Fundamental limitations for quantum and nanoscale thermodynamics, Nature Comm. 4, 2059 (2013).
\bibitem{Farin} D. Farina, G. M. Andolina, A. Mari, M. Polini and V. Giovannetti, powerful charging of quantum batteries, Phys. Rev. B 99, 035421 (2019).
\bibitem{Zhang} Y-Y. Zhang, T-R. Yang, L. Fu and X. Wang, Powerful harmonic charging in a quantum battery, Phys. Rev. E 99, 052106 (2019).
\bibitem{Fus} L. Fusco, M. Paternostro, and G. D. Chiara, Work extraction and energy storage in the Dicke model, Phys. Rev. E 94, 052122 (2016).
\bibitem{Cata} R. R. Rodriguez, B. Ahmadi, P. Mazurek, S. Barzanjeh, R. Alicki and P. Horodecki, catalysis in charging quantum batteries, Phys. Rev. A 107, 042419 (2023).
\bibitem{Maze} J. Carrasco, J. R. Maze, C. Hermann-Avigliano and F. Barra, collective enhancement in dissipative quantum batteries, Phys. Rev. E. 105, 064119 (2022).
\bibitem{Manzo} M. Gumberidze, M. Kol\'{a}r and R. filip, Measurement induced Synthesis of coherent Quantum Batteries, Sci. Rep 9, 19628 (2019).
\bibitem{Cond} D. Ferraro, M. Campisi, G. M. Andolina, V. Pellegrini and M. Polini, High-power collective charging of a solid-state quantum battery, Phys. Rev. Lett. 120, 117702 (2018).
\bibitem{Forn} P. Forn-D\'{\i}laz, J. J. Garc\'{\i}la-Ripoll, B. Peropadre, J.-L. Orgiazzi, M. A. Yurtalan, R. Belyansky, C. M. Wilson, and A. Lupascu, Ultrastrong coupling of a single artificial atom to an electromagnetic continuum in the nonperturbative regime, Nat. Phys. 13, 39 (2016).
\bibitem{Lv} Bruzewicz, C.D.; Chiaverini, J.; McConnell, R.; Sage, J.M. Trapped-Ion Quantum Computing: Progress and Challenges. Appl. Phys. Rev. 2019, 6, 021314..
\bibitem{Bau} K. Baumann, C. Guerlin, F. Brennecke, and T. Esslinger, The dicke quantum phase transition with a superfluid gas in an optical cavity, Nature (London) 464, 1301 (2010)
\bibitem{Devoret} Devoret, M.H.; Schoelkopf, R. J. Superconducting Circuits for Quantum Information: An Outlook. Science 2013, 339, 1169
\bibitem{Farin1} D. Farina, G. M. Andolina, A. Mari, M. Polini, and V. Giovannetti, Charger-mediated energy transfer for quantum batteries: Anopen-system approach. Phys. Rev. B 99, 035421 (2019).
\bibitem{Camp} C. Ou, R. V. Chamberlin and S. Abe, Lindbladian operators, von Neumann entropy and energy conservation in time-dependent quantum open systems, Physica A 466, 450 (2017).
\bibitem{Carega} M. Carrega, A. Crescente, D. Ferraro, and M. Sassetti, Dissipative dynamics of an open quantum battery. New J. Phys. 22, 083085 (2020).
\bibitem{Barra} F. Barra, Dissipative charging of a quantum battery, Phys. Rev. Lett. 122, 210601 (2019).
\bibitem{San0} A. C. Santos, Quantum advantage of two-level batteries in
self-discharging process, Phys. Rev. E 103, 042118 (2021).
\bibitem{Pedro} L. P. Garcia-Pintos, A. Hamma, A. del Campo, Fluctuations in extractable work bound the charging power of quantum batteries. Phys. Rev. Lett. 125, 040601 (2020).
\bibitem{Salimi} F. H. Kamian, F. T. Tabesh, S. Salimi, F. Kheirandish, and A. C. Santos, Non-markovian effects on charging and selfdischarging processes of quantum batteries, New J. Phys. 22, 083007 (2020).
\bibitem{Kamin1} F. T. Tabesh, F. H. Kamin, and S. Salimi, Environmentmediated charging process of quantum batteries, Phys. Rev. A 102, 052223 (2020).
\bibitem{Squeezing} F. Centrone, L. Mancino, M. Paternostro, Charging batteries with quantum squeezing, https://doi.org/10.48550/arXiv.2106.07899.
\bibitem{Dark} J. Q. Quach and W. J. Munro, Using dark states to charge and stabilise open quantum batteries, Phys. Rev. Applied 14, 024092 (2020).
\bibitem{Mitch} M. T. Mitchison, J. Goold and J. Prior, Charging a quantum battery with linear feedback control, Quantum 5, 500 (2021).
\bibitem{Shao} Y. Yao and X. Q. Shao, Phys. Rev. E Optimal charging of open spin-chain quantum batteries via homodyne-based feedback control, 106, 014138 (2022).
\bibitem{Ios} S. Borisenok, Ergotropy of quantum battery controlled via target attractor feedback, J. Appl. Phys. 12, 43 (2020).
\bibitem {Borhan} R. R. Rodriguez, B. Ahmadi, G. Suarez, P. Mazurek, S. Barzanjeh, P. Horodecki, Optimal quantum control of charging quantum batteries, arXiv:2207.00094 [quant-ph].
\bibitem{Franc} F. Mazzoncini, V. Cavina, G. M. Andolina, P. A. Erdman and V. Giovannetti, Optimal control methods for quantum batteries, Phys. Rev. A 107 (2023) 032218.
\bibitem{Behzadi} N. Behzadi and H. Kassani, Mechanism of controlling robust and stable charging of open quantum batteries, J. Phys. A: Math. Theor. 55, 425303 (2022).
\bibitem{Yu0} J. L. Li, H. Z. Shen and X. X. Yi, Quantum batteries in non-Markovian reservoirs, Opt. Lett 21, 5614 (2022).
\bibitem{Baris} A. C. Santos, B. \c{C}akmak, S. Campbell and N.T. Zinner, Stable adiabatic quantum batteries, Phys. Rev. E 100, 032107 (2019).
\bibitem{Segal} J. Liu, D. Segal, Boosting quantum battery performance by structure engineering, arXiv:2104.06522 [quant-ph].
\bibitem{Epjp0} J. Taghipour, B. Mojaveri and A. Dehghani, Witnessing entanglement between two two-level atoms coupled to a leaky cavity via two-photon relaxation, Eur. Phys. J. Plus 137, 772 (2022).
\bibitem{morteza0} A. Mortezapour, M. A. Borji, and R. L. Franco, Protecting entanglement by adjusting the velocities of moving qubits inside non-Markovian environments, Laser Phys. Lett 14, 055201 (2017).
\bibitem{Chao0} W. Chao and F. Mao-Fa, The entanglement of two moving atoms interacting with a single-mode field via a three-photon process, Chin. Phys. B 19, 020309 (2010).
\bibitem{sare0} S. Golkar and M. K. Tavassoly and A. Nourmandipour, Entanglement dynamics of moving qubits in a common environment, J. Opt. Soc. Am. B 37, 400 (2020).
\bibitem{Golkar1} S. Golkar and M. K. Tavassoly And A. Nourmandipour, Qubit movement-assisted entanglement swapping, Chin. Phys. B. 29, 050304 (2020).
\bibitem{Epjp1} B. Mojaveri, A. Dehghani and J. Taghipour, Control of entanglement, single excited-state population and memory-assisted entropic uncertainty of two qubits moving in a cavity by using a classical driving field, Eur. Phys. J. Plus 137, 1065 (2022).
\bibitem{MPLA} J. Taghipour, B. Mojaveri and A. Dehghani, Witnessing entanglement between two two-level atoms moving inside a leaky cavity under classical control, Mod. Phys. Lett. A 37, 2250141 (2022).
\bibitem{Wang00} Q. Wang, R. Liu, H. M. Zou, D. Long and J. Wang, Entanglement dynamics of an open moving-biparticle system driven by classical-field, Phys. Scr. 97, 055101, (2022).
\bibitem{Allahverdyan} A. E. Allahverdyan, R. Balian and T. M. Nieuwenhuizen, Maximal work extraction from finite quantum systems. Eur. phys. Lett 67, 565 (2004).
\bibitem{Franc0} G. Francica, J. Goold, F. Plastina, and M. Paternostro, Daemonic ergotropy: enhanced work extraction from quantum correlations, npj Quantum Inf. 3, 12 (2017).
\bibitem{Cakmak0} B. \c{C}akmak, Ergotropy from coherences in an open quantum system, Phys. Rev. E 102, 042111 (2020).
\bibitem{Engl} B.G. Englert, J. Schwinger, A.O. Barut and M.O. Scully, Reflecting slow atoms from a micromaser field, Eur. Phys. Lett 14, 25 (1991).
\bibitem{Haro} S. Haroche, M. Brune and J.M. Raimond, Trapping atoms by the vacuum field in a cavity, Eur. Phys. Lett 14, 19 (1991).
\bibitem{Lenard} C. Leonardi and A. Vagliea, Non-markovian dynamics and spectrum of a moving atom strongly coupled to the field in a damped cavity, Opt. Commun 97, 130 (1993).
\bibitem{mortezapour} F. Nosrati, A. Mortezapour and R. Lo Franco, Validating and controlling quantum enhancement against noise by the motion of a qubit, Phys. Rev. A. 101, 012331 (2020).
\bibitem{Cook} R. J. Cook, Atomic motion in resonant radiation: An application of Ehrenfest's theorem, Phys. Rev. A. 20, 224 (1979).
\bibitem{Wilkens} M. Wilkens, Z. Bialynicka-Birula and P. Meystre, Spontaneous emission in a Fabry-P\'{e}rot cavity: The effects of atomic motion, Phys. Rev. A. 45, 477 (1992).
\bibitem{Breuer0} H. P. Breuer and F. Petruccione, \textit{The Theory of Open Quantum Systems} (Oxford University Press, Oxford, New York, 2002).
\bibitem{Hood} C. J. Hood et al., The Atom-Cavity Microscope: Single Atoms Bound in Orbit by Single Photons, Science 287, 1447 (2000).
\bibitem{Pinkse} P. W. H. Pinkse et al., Trapping an atom with single photons, Nature 404, 365 (2000).
\bibitem{Jones} P. J. Jones, J. A. M. Huhtam\"{a}ki, K. Y. Tan and M. M\"{o}tt\"{o}nen, Tunable electromagnetic environment for superconducting quantum bits, Sci. Rep. 3, 1987 (2013).
\end{thebibliography}
\end{document}